\documentclass[preprint,aps,tighten]{revtex4}

\usepackage{graphicx}
\begin{document}

\title{ Pion-Muon Asymmetry Revisited }

\author{W. A. Perkins}

\affiliation {Perkins Advanced Computing Systems,\\ 12303 Hidden Meadows
Circle, Auburn, CA 95603, USA\\E-mail: wperkins@nccn.net} 

\begin{abstract}
Long ago an unexpected and unexplainable phenomena was observed. The distribution of muons from positive pion decay at rest was anisotropic with an excess in the backward direction relative to the direction of the proton beam from which the pions were created. Although this effect was observed by several different groups with pions produced by different means, the result was not accepted by the physics community, because it is in direct conflict with a large set of other experiments indicating that the pion is a pseudoscalar particle. It is possible to satisfy both sets of experiments if helicity-zero vector particles exist and the pion is such a particle. Helicity-zero vector particles have direction but no net spin. For the neutral pion to be a vector particle requires an additional modification to conventional theory as discussed herein. An experiment is proposed which can prove that the asymmetry in the distribution of muons from pion decay is a genuine physical effect because the asymmetry can be modified in a controllable manner. A positive result will also prove that the pion is {\it not} a pseudoscalar particle. 

\end{abstract}




\maketitle

\section{\label{sec.intro}Introduction}

First of all, we will discuss in some detail the experimental evidence indicating that pions have an internal direction. After that, we consider a new model for the pion that can accommodate both this result and the evidence indicating that the pion is a pseudoscalar.
 
Fifty years ago after parity non-conservation was discovered in the $\pi^+ \rightarrow \mu^+ \rightarrow e^+$ chain with an anisotropic angular distribution of electrons from muon decay, experimentalists looked at the distribution of muons from pion decay. (In some cases this was done to check that their method would show an isotropic distribution for a pseudoscalar particle.) To their surprise they found a large anisotropic angular distribution of muons from positive pion decay at rest. 

The history of physics is full of observed phenomena that were later retracted or shown to be invalid by other scientists. However, there are several reasons why this observation (anisotropic angular distribution of muons from pion decay at rest) looks like a genuine physical effect: (1) It was observed by several different groups~\cite{hulubei1,peterson,bruin,lattes,osborne,ammar,bhowmik}, even in one experiment~\cite{peterson} that predated the observation of the anisotropic muon-electron angular distribution, (2) It was observed in pions produced by different methods (proton-proton interactions and kaon decay), (3) It was observed with repeatable results by one group~\cite{hulubei1} over many years, and (4) It was almost always detected as a forward-backward asymmetry relative to the proton beam direction with a surplus in the backward direction, although sometimes an equator-pole asymmetry was observed.

Several other experimental groups~\cite{garwin1,ferretti,bogachev,vaisenberg} obtained results that showed appreciable asymmetry in the muon distributions, but, in summary, they reported that the distributions were isotropic. Most of these negative findings were re-examined by Hulubei~{\it et.~al.}~\cite{hulubei1} showing that ``several authors, yielding to general opinion, have formulated negative conclusions in spite of their own positive results.'' 

With regard to the counter experiments of Garwin~{\it et.~al.}~\cite{garwin1}, the direction of pions creation was in the x-y plane. They measured the muon asymmetry along x, y, and z axes, detecting small asymmetries along the y and z axes and none along the x axis. If one assumes that the pions have longitudinal polarization along or near the y-axis, their results indicate a measurable forward-backward asymmetry and equator-pole asymmetry. One would expect a smaller effect here because they are using the $\mu-e$ decay asymmetry to detect the muon asymmetry. 

The electronic counter experiments of Crewe~{\it et.~al.}~\cite{crewe} were performed with an identical experimental arrangement to that used in the emulsion exposure of Lattes~\cite{lattes}. The pions were produced inside the cyclotron and extracted at 180 degrees to the direction of the proton beam.
Interestingly, in both experiments the pions were bent by magnetic fields such that they entered the detectors at an angle very close to the direction at which the proton beam struck the target to produce the pions. Lattes~\cite{lattes} had found a distribution of muons from pion decay that deviated significantly 
from isotropy with a longitudinal parameter, $a = -.070 \pm .018$ and a latitude parameter $b = -.106 \pm .018$. (In other words, they observed a forward-backward asymmetry and an equator-pole asymmetry.) In the counter 
experiment~\cite{crewe}, they looked for transverse polarization by measuring  
any asymmetry in left-right and up-down scattering and $\pi-\mu$ decay in flight. They did not find any asymmetry, but this is to be expected because they only looked for a transverse polarization while the polarization observed by Lattes~\cite{lattes} under the same conditions was longitudinal. A modified version of this counter experiment in which the pions are bent magnetically such that their momentum is at an angle of 90 degrees to the proton-beam direction (rather than parallel to it) is discussed in Sec.~\ref{sec.exper}. 

In a third group of 
experiments~\cite{hulubei2,alston,alikhanian,balandin,barmin,abashian,%
connolly,taylor,gbc,frota1,frota2} the distributions of muons were observed to be isotropic. In order to show that using their techniques would result in an isotropic muon distribution if the pions were not polarized, in 1965, Hulubei~{\it et.~al.}~\cite{hulubei2} performed the experiment again under radically different conditions. This resulted in an isotropic muon distribution, very different from the earlier one.

Lack of significant asymmetric muon distributions in the bubble chamber experiments~\cite{alston,alikhanian,balandin,barmin,abashian} may have resulted because ``No systematic measurements have made in order to separate beam muon from straight $\pi-\mu$ decays,'' as noted by Hulubei~{\it et.~al.} Since the pions of Ref.~\cite{abashian} resulted from the bombardment of a copper target instead of a hydrogen-rich target, they are likely to be unpolarized.
A possible explanation for the observed isotropic muon distributions in Ref.~\cite{connolly} may have been lack of pion polarization, since they studied photoproduced pions.
 
The observed isotropic muon distributions in Ref.~\cite{taylor,gbc} using pions from kaon decay is puzzling since an asymmetry was observed in the similar, earlier experiments of Ref.~\cite{bruin,garwin1}. Osborne~\cite{osborne} found very large asymmetries in his kaon experiments (e.g., distributions containing 19 positive versus 61 negative members) by choosing a polarization axis that was dependent upon the direction of the proton beam that created the kaon.  

In 1969, Frota-Pessoa~\cite{frota1} reanalyzed the emulsion stacks used in the experiments of Hulubei {\it et.~al.}. The apparent goal of this work was to show that the anomaly (anisotropic muon distribution) did not really exist. It is not surprising that he concluded, ``The results of two experiments, which seems to be the safe part of scanning of this stack, are in good agreement with isotropy.'' Hulubei~{\it et.~al.}~\cite{hulubei1} commented on an earlier work by Frota-Pessoa and Margem~\cite{frota2} in that regard, ``The experiment was performed on plates from our stack and would therefore seem to be most appropriate for a comparison. Unfortunately, the conclusions drawn in that paper are based on qualitative considerations.''

The chance of a statistical fluctuation causing the observe asymmetry is one in 2500 in just one set of experiments~\cite{hulubei1} and much less when all the experiments are considered. Hulubei~{\it et.~al.}~\cite{hulubei1} have shown that systematic errors could not have caused the effect. Lattes~\cite{lattes} examined seven possible sources of error and showed that none of then could account for the observed effect.

In attempting to explain the asymmetric muon distributions in the 1960's, 
scientists~\cite{gbc2,weiner,banyai} suggested that the effect was caused by a new particle (present in the pion beam) with spin but mass degenerate with that of the pion. These attempts were not successful since a spin-1 pion should decay through the electron mode as often as the muon mode and because no other evidence for such pion-like particles was ever found. Cassels~\cite{cassels} noted that other alternatives besides the pion having spin were possible.

How can the many experiments~\cite{durbin,clark,cartwright} showing that pions have spin-0 and the observations~\cite{hulubei1,peterson,bruin,lattes,osborne,ammar,bhowmik} that pions have a direction be compatible? In a recent paper~\cite{perkins1} the author has proposed the existence of helicity-zero particles, and the pion could be such a particle. As formulated, helicity-zero vector particles are similar to pseudoscalar particles because they have no net spin. Whereas the polarization of a spin-1 vector particle is defined by sixteen or more parameters~\cite{titeica}, an helicity-zero particle has only longitudinal polarization which agrees with the observed polarization. Since it has zero spin, an helicity-zero vector pion satisfies experimental results such as the large ratio of the $\pi-\mu$ decay mode relative to the $\pi-e$ decay mode~\cite{rinaudo,taylor} and the observed polarization of muons from pion decay~\cite{garwin2}. 

Experiments~\cite{carrigan} have shown that negative pions have zero (or very small) magnetic moment. It was noted in one of the asymmetry experiments~\cite{peterson}, that the asymmetry existed in a high magnetic field. Since pion precession due to a magnetic moment should have washed out the observed asymmetry, this experiment indicates small or zero magnetic moment for the positive pions causing the asymmetry. An helicity-zero vector pion would have zero magnetic moment because it has zero net spin.

If charged pions are helicity-zero vector particles, the neutral pion must also be an helicity-zero vector particle. The well-known proofs of Landau~\cite{landau} and Yang~\cite{yang} showing that the neutral pion cannot be a vector particle are based on the assumption that a state of two photons must be symmetric under interchange of the photons. Although this assumption seems very reasonable, as discussed in Sec.~6.2 of Ref.~\cite{perkins2}, a state of two {\it composite} photons need not be symmetric under interchange if the two photons are not identical.

\section{\label{sec.exper}Experimental Test}

Using a vector-meson model for the pion, Wentzel~\cite{wentzel} showed (for longitudinally polarized pions) that the polarization would be nearly complete in the direction of the producing nucleon beam.
We have tried to determine the pion polarization direction from the asymmetry experiments. The asymmetry results are reported relative to an axis of the emulsion plates, which are usually aligned with the pion beam and in the plane of pion creation by the proton beam. The direction of the pion beam at the emulsions is usually within 25 degrees of its direction at creation. In most of the experiments~\cite{hulubei1,peterson,lattes,bogachev,vaisenberg} the asymmetry appears as a forward-backward deviation with an excess in the backward direction relative to the proton beam direction. An equator-pole or latitude asymmetry was observed in some experiments~\cite{hulubei1,lattes,ammar,ferretti}. 

Unlike the previous $\pi-\mu$ asymmetry experiments, we propose two experiments that involve the variation of a parameter to change the asymmetry in a controllable manner. 
As noted above, the pions created in proton-proton interactions tend to be longitudinally polarized in the proton-beam direction with an excess of muons decays in the backward direction. In order to obtain a polarized beam of pions, Hulubei {\it et.~al.} recommended that the pions should be created outside the accelerator field from a proton beam striking a hydrogen or hydrogen-rich target so that the pions are created in two-particle reactions. Furthermore, the pions should be extracted in the forward direction. The experiments of Lattes~\cite{lattes}, however, show that polarized pions can be obtained from protons striking a beryllium target inside a cyclotron and extracted at 180 degrees to the proton beam. As discussed in Sec.~\ref{sec.intro}, pions have zero magnetic moment. Thus their polarization direction after creation will not be affected by a magnetic field.


\begin{figure}
\includegraphics[scale=.5]{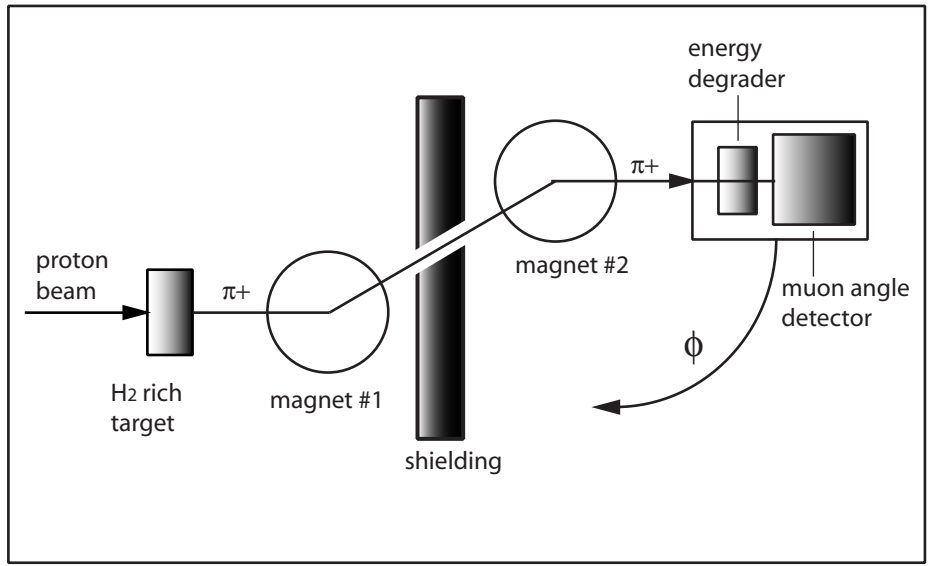}
\caption{Illustration of an experimental apparatus for detecting $\pi-\mu$ asymmetry. The detection box should supported is such a way that it can be rotated in a circular arc about the bending axis of the second magnet.}
\label{f1}
\end{figure}

One experimental test involves varying the angle at which the pions enter the detector (relative to their creation direction) and measuring the angle of peak muon emission. Figure~\ref{f1} indicates how this might be accomplished. The first magnet before the shielding is used to select the pion energy. The second magnet varies the pion angle and the detector is rotated accordingly. The energy degrader slows the pions so that they will come to rest in the muon-angle detector, which could be emulsions, for example.

\begin{figure}
\includegraphics[scale=.5]{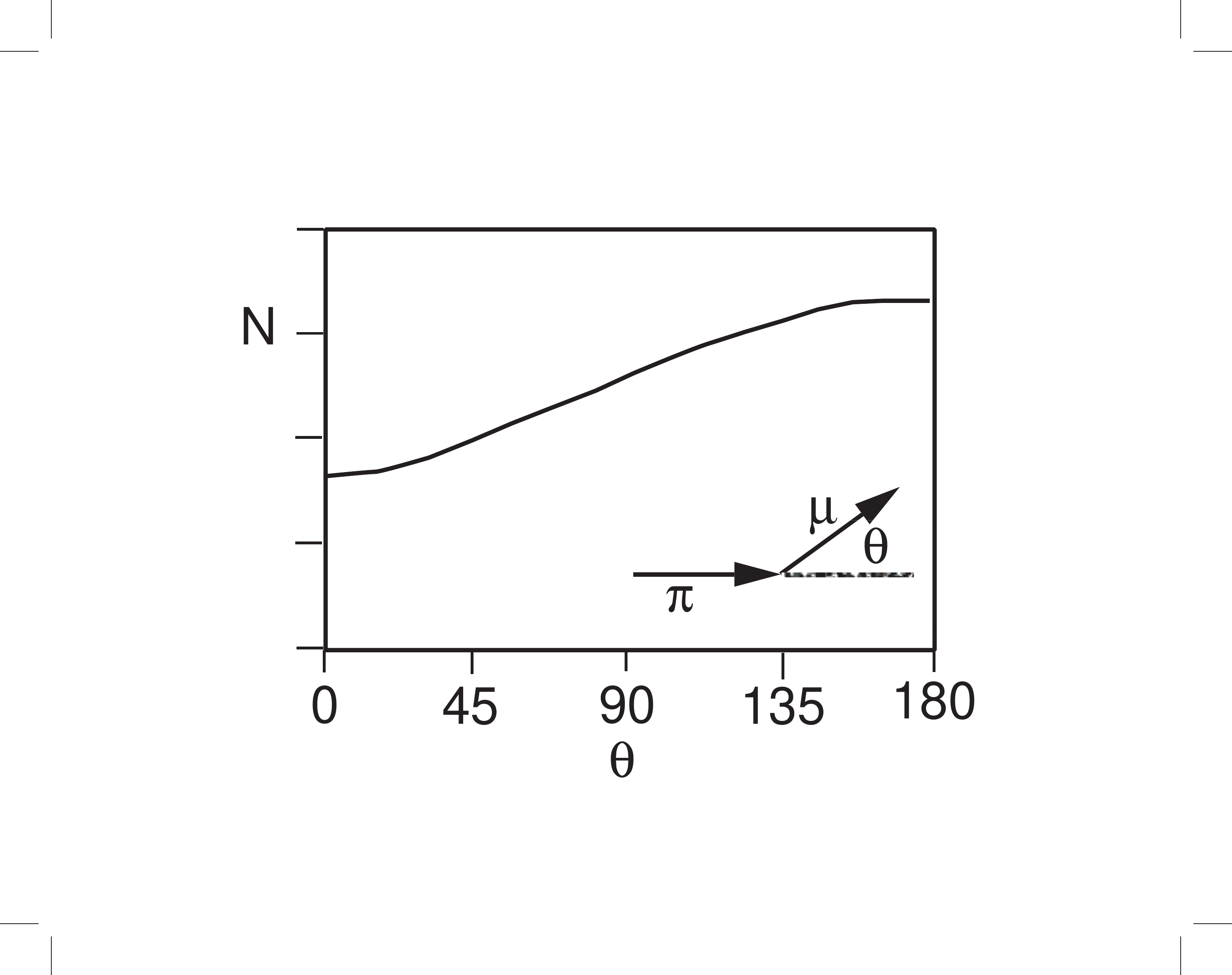}
\caption{Expected angular distribution of muons from pion decay. Peak is at 180 degrees from the proton-beam direction (and pion-beam direction for $\phi$ = 0).}
\label{f2}
\end{figure}

\begin{figure}
\includegraphics[scale=.5]{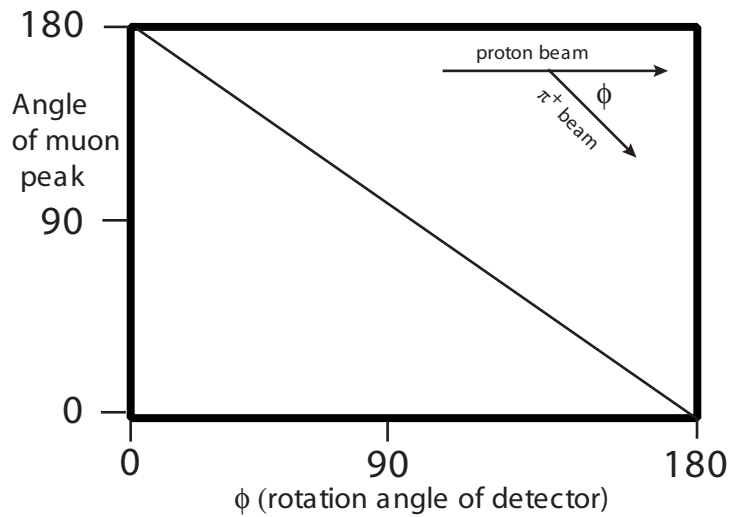}
\caption{Expected variation in the angle at which the peak in the muon distribution occurs (as measured relative to pion momentum) as a function of the rotation angle of the detector.}
\label{f3}
\end{figure}

With the pion angle set as shown in the diagram, we expect that the detector will record an anisotropic muon distribution with a peak in the backward direction as shown in Fig.~\ref{f2}. This is essentially the experiment of 
Hulubei {\it et.~al.}~\cite{hulubei1}. Since the pion-polarization direction is not changed by the magnetic field, varying the angle of pion momentum with the second magnet and rotating the detector thorough an angle $\phi$ will cause the angle of the muon distribution peak to change as shown in Fig.~\ref{f3}. A result, similar to that in Fig.~\ref{f3}, will be obtained by rotating the detector thorough an angle of $-\phi$. However, the polarization direction for $\phi$ = -90 degrees will be opposite to that for $\phi$ = 90 degrees, and can provide further evidence of a controllable asymmetry.


A second experimental test involves scattering and decay of pions in flight. This is essentially the counter experiment of 
Crewe {\it et.~al.}~\cite{crewe} with one importance difference. The pion beam should be magnetically bent by 90 degrees so that its direction is perpendicular to the axis of the proton beam at pion creation. This will result in the polarization direction of the pions being transverse to their momentum. Scattering the pions as in the counter experiment~\cite{crewe} should result in a left-right asymmetry. Bending the pions through an angle of 90 degrees in the opposite direction should reverse the left-right asymmetry. One can also look for an asymmetry in the $\pi-\mu$ decay in flight as in the experiment of Crewe {\it et.~al.}~\cite{crewe}, but the polarization axis must be perpendicular to the momentum of the pions.

\section{\label{sec.concl}Conclusion }
   
Over the past forty years no experiment has been performed that explained away the observed $\pi-\mu$ 
asymmetry~\cite{hulubei1,peterson,bruin,lattes,osborne,ammar,%
bhowmik,garwin1,ferretti,bogachev,vaisenberg}. If the $\pi-\mu$ asymmetry results~\cite{hulubei1,peterson,bruin,lattes,osborne,ammar,%
bhowmik,garwin1,ferretti,bogachev,vaisenberg} are not interpreted as evidence of non-zero pion spin, then they are not in conflict with the many experiments showing that the pion has spin $0$~\cite{durbin,clark,cartwright,rinaudo,taylor,garwin2,carrigan}.
Furthermore, two experiments in which the asymmetry axis is modified in a controlled manner have been proposed. A positive result from either of these experiment will prove that the pion is {\it not} a pseudoscalar particle. We recommend the first experiment because it has an extremely high probability of giving a positive result, based on the experiments of 
Hulubei {\it et.~al.}~\cite{hulubei1}. However, the scattering experiment may be simpler to perform.

It will be necessary to reconcile a positive result in this $\pi-\mu$ asymmetry experiment with our existing pion knowledge. Although we have pointed out that the main features of the pion can be handled by an helicity-zero-particle model~\cite{perkins1}, a systematic study is obviously needed. 

Yukawa~\cite{yukawa} and others expected that the pion would be a {\it vector} particle, before early experiments showed that the pion had spin zero and odd parity which were interpreted to mean that the pion is a pseudoscalar. In 1960, 
Sakurai~\cite{sakurai} noted that the ``global symmetry'' model of Gell-Mann and Schwinger failed because ``unfortunately, the pion field is pseudoscalar not vector.'' It is also interesting that experimentally~\cite{lafferty}, ``light vector mesons have been found to populate preferentially in the helicity-zero state.'' 
                                  
\acknowledgments

This work was inspired by the experimental results of the Romanian Group (H.~Hulubei, J.~S.~Auslander,  E.~M.~Friedlander, and S.~Titeica).
One has to admire the courage of that Group for their determination to stand by their experimental results over many years in spite of the natural desire of the physics community to resolve or eliminate the anomaly. To satisfy objections to their results, they performed more experiments and made numerous additional checks for systematic errors. In 1965, they even performed the experiment again to show that their method would give an isotropic distribution if the pion beam were unpolarized~\cite{hulubei2}.

\end{document}